\documentclass[]{gAPA2e}

\newcommand{\dsp}{\displaystyle}

\newcommand{\LL}{\mathcal{L}}
\newcommand{\MM}{\mathcal{M}}
\newcommand{\LH}{\tilde{\LL}}
\newcommand{\MH}{\tilde{\MM}}
\newcommand{\XX}{\mathbf{X}}
\newcommand{\TT}{\mathbf{T}}
\newcommand{\XXH}{\hat{\XX}} 
\newcommand{\TTH}{\hat{\TT}} 

\newcommand{\myG}{\mathbf{G}}
\newcommand{\zmat}{\mathbf{0}}

\newcommand{\DD}{\mathsf{D}}
\newcommand{\II}{\mathsf{I}}
\newcommand{\opz}{\mathsf{O}}
\newcommand{\eqe}{\doteq}
\newcommand{\pvec}{\Psi}

\newcommand{\lax}{Lax pair}
\newcommand{\laxeq}{Lax equation}

\newcommand{\kdv}{KdV}

\usepackage{xcolor}
\usepackage{verbatim}

\newcommand{\bq}{\begin{equation}}
\newcommand{\eq}{\end{equation}}

\newcommand{\tf}[2][1]{\tfrac{#1}{#2}}
\newcommand{\mat}[2][*\c@MaxMatrixCols c]{\left( \hskip - \arraycolsep 
\array{#1} #2 \endarray \hskip - \arraycolsep \right)}
\newcommand{\mlabel}[1]{\\[-0.5\baselineskip] &  \label{#1} \\[-0.5\baselineskip]}
\newcommand{\gS}[1][\relax]{\ensuremath{\mathcal{S}^{#1}}}
\newcommand{\cov}[1][x]{\ensuremath{\nabla_{#1}}}

\DeclareSymbolFont{AMSb}{U}{msb}{m}{n}
\DeclareMathSymbol{\Real}{\mathalpha}{AMSb}{"52}

\begin{document}

\doi{10.1080/0003681YYxxxxxxxx}
\issn{1563-504X}
\issnp{0003-6811}
\jvol{00} \jnum{00} \jyear{2009} \jmonth{January}


\markboth{M.\ Hickman et al.}{Applicable Analysis}

\articletype{}

\title{{\sc Dedicated to the memory of Alan Jeffrey} \\[5mm] 
Scaling invariant Lax pairs of nonlinear evolution equations}

\author{Mark Hickman$^{a\ast}$, Willy Hereman$^{b}$\thanks
        {$^\ast$Corresponding author. Email: mark.hickman@canterbury.ac.nz}, 
        Jennifer Larue$^b$ and \"{U}nal G\"{o}kta\c{s}$^c$
        \\ \vspace{6pt}  
        \em{$^a$Department of Mathematics and Statistics,\\ 
        University of Canterbury, Christchurch, New Zealand; \\
        $^b$Department of Applied Mathematics and Statistics,\\ 
        Colorado School of Mines, Golden, CO, USA; \\
        $^c$Department of Computer Engineering, \\
        Turgut \"{O}zal University, Ke\c{c}i\"{o}ren, Ankara, Turkey} 
       }

\maketitle

\begin{abstract}
A completely integrable nonlinear partial differential equation (PDE) 
can be associated with a system of linear PDEs in an auxiliary function 
whose compatibility requires that the original PDE is satisfied.
This associated system is called a Lax pair.
Two equivalent representations are presented.
The first uses a pair of differential operators which leads to a higher order 
%
%
linear system for the auxiliary function.
The second uses a pair of matrices which leads to a first-order linear system. 
In this paper we present a method, which is easily implemented in {\sc Maple} 
or {\sc Mathematica}, to compute an operator Lax pair for a set of PDEs.
 
In the operator representation, the determining equations for the Lax pair 
split into a set of kinematic constraints which are independent of the 
original equation and a set of dynamical equations which do depend on it. 
The kinematic constraints can be solved generically. 
We assume that the operators have a scaling symmetry. 
The dynamical equations are then reduced to a set of nonlinear algebraic 
equations. 

This approach is illustrated with well-known examples from soliton theory. 
In particular, it is applied to a three parameter class of fifth-order 
KdV-like evolution equations which includes the Lax fifth-order KdV, 
Sawada-Kotera and Kaup-Kuperschmidt equations.
A second Lax pair was found for the Sawada--Kotera equation. 
\begin{classcode}
Primary: 37J35, 37K40, 35Q51; 
Secondary: 68W30, 47J35, 70H06.
\end{classcode}
\end{abstract}
\vspace{-5mm}
\noindent
\section{Introduction}
\label{sec-introduction}
The complete integrability of a system of partial differential equations 
has been a topic of active research over the last forty or so years. 
Even today there is no one generally accepted definition of integrability. 
Many approaches have been advocated (see \cite{MikhailovLNP09} for a recent 
review), all of which have their merits but none seem to encapsulate the 
essence of integrability. 
However one concept has appeared in many different approaches; the Lax pair. 
This is a reformulation of the original system of nonlinear equations as 
the compatibility condition for a system of {\em linear} equations. 

The story of Lax pairs begins with the discovery by Lax \cite{Lax1968} 
of such a linear system for the ubiquitous KdV equation. 
This equation \cite{Bous1877,KdV1895} models a variety of nonlinear wave 
phenomena, including shallow water waves \cite{WHencyclop2008} and 
ion-acoustic waves in plasmas 
\cite{MAandPCbook1991,MAandHSbook1981,PDandRJbook1989}. 
Lax's discovery shed light on the then newly discovered inverse scattering 
transform method \cite{CGandJGandMKandRM1967} to solve the KdV equation. 
It allowed this method to be extended to a variety of integrable equations. 
In 1979 it was realised that the Lax pair could be interpreted as a 
zero curvature condition on an appropriate connection \cite{ZakharovFAA79}. 
This led to generalisations of the method of inverse scattering transforms.
In the 1980s, the algebraic structure of Lax pairs was elucidated. 
The connection between Lax pairs and Kac-Moody algebras was given in 
\cite{VDandVSjsm1984}. 
Subsequent applications for Lax pairs include B\"{a}cklund-Darboux 
transformations \cite{Rogers02}, recursion operators \cite{MGetal1999} and 
generating integrable hierarchies via the root method 
\cite{GelfandDikiiFAA76b,OhtaPTP88,YJandYC2008}. 
However there was a problem; finding the Lax pair in the first place. 

A Lax pair is associated with an infinite hierarchy of local conservation 
laws; a harbinger of complete integrability (though not all equations 
with a Lax pair have an infinite set of conservation laws). 
Exploiting this connection, Wahlquist and Estabrook 
\cite{HWandFE1975b,HWandFE1975a} proposed a method based on pseudopotentials 
that, in certain circumstances, leads to Lax pairs. 
However their method leads to a nontrivial problem of finding a 
representation of a Lie algebra when only a subset of the commutation 
relations are known.

Another approach to the construction of a Lax pair is provided by 
singularity analysis. 
In 1977 it was noted \cite{AblowitzPRL77} that all symmetry reductions of 
the classical completely integrable equations result in equations that can 
be transformed to Painlev\'{e} equations. 
This observation subsequently gave rise to the so-called Painlev\'{e} test 
%
%
\cite{JWandMTandGC1983}. 
In this approach Lax pairs are generated from truncated Painlev\'{e} 
expansions \cite{MMandRC1991}.
However the complexity of Painlev\'{e} expansion depends critically on the 
choice of expansion variable. 

Computer algebra packages have also been employed to symbolically verify 
Lax pairs \cite{MI1985,JL2011}. 
Of course this requires prior knowledge (or a good guess) of the Lax pair. 

In this paper we address the issue of finding Lax pairs, in operator form, 
for a given system of differential equations by an approach that is 
amenable to computer algebra. 
The determining equations for the Lax pair is split into two sets. 
The first set, the kinematic constraints, are solved generically. 
The second set, the dynamical equations, depend on the system under 
consideration. 
We make the assumption that the system has a scaling symmetry and that this 
symmetry is inherited by the Lax pair. 
This allows us to reduce the dynamical equations to an overdetermined 
system of algebraic equations (nonlinear, naturally). 
Such systems may then be solved by Gr\"{o}bner basis techniques. 

In the next section, Lax pairs are introduced in their operator 
representation. 
In Section~\ref{sec-laxpairsmatrices} the matrix representation of Lax pairs 
is discussed. 
The relationship between these two representations is also examined. 
Section~\ref{dilationinvariancePDEs} introduces the concept of scaling 
symmetries. 
The algorithm to compute Lax pairs based upon scaling symmetries is outlined 
in the next section. 
Examples of the algorithm applied to some well-known equations are given 
in Section~\ref{sec-examples}. 
Finally the algorithm is used to classify the integrable subcases of a 
fifth--order KdV-like equation with three parameters. 
A second Lax pair was found for the Sawada--Kotera equation. 

\section{Lax Pairs in Operator Form}
\label{sec-laxpairsoperators}
In this paper we consider nonlinear systems of {\it evolution} equations 
in $(1+1)$ dimensions, 
\begin{equation}
\label{systempde}
{\bf u}_t = {\bf F}({\bf u}, {\bf u}_x, {\bf u}_{2x}, \dots ),
\end{equation}
where $x$ and $t$ are the space and time variables, respectively.
The vector ${\bf u}(x,t)$ has $N$ components $u_i$ and ${\bf F}$ is a 
nonlinear function of its arguments.
In the examples we denote the components of ${\bf u}$ by $u,v,w,\ldots$.
Throughout the paper we use the subscript notation for partial derivatives.
If parameters are present in (\ref{systempde}), they will be denoted by 
lower-case Greek letters.

%
In his seminal paper \cite{Lax1968}, Lax showed that completely integrable 
{\it nonlinear} PDEs have an associated system of {\it linear} PDEs in an 
auxiliary function $\psi(x,t),$ 
\begin{align}
\LL \psi &= \lambda \psi,  \notag
\mlabel{LaxOpEq}  
\psi_t &= \MM \psi  \notag  
\end{align}
%
where $\LL$ and $\MM$ are linear differential operators
(expressed in powers of the total derivative operator $\DD_x$ for the space 
variable $x).$ 
$\psi$ is an eigenfunction of $\LL$ corresponding to eigenvalue $\lambda.$ 
The operators $(\LL,\, \MM)$ are now known as a {\em Lax pair} for  
\eqref{systempde}.
The property that \eqref{systempde} is completely integrable is reflected 
in the fact that the eigenvalues do not change with time which makes the 
problem {\em isospectral}.

Let $G$ be a differential function (functional); that is, a function of 
$x, t, u$ and partial derivatives of $u$ and let 
$u_{p,q} = \partial_x^p \, \partial_t^q \,u$. 
The {\em total derivatives} of $G$ is given by
\begin{align}
\DD_x G &= \frac{\partial G}{\partial x} +
\sum_{p,q} \frac{\partial G}{\partial u_{p,q}} u_{p+1,q}, \notag
\mlabel{TotalD}
\DD_t G &= \frac{\partial G}{\partial t} +
\sum_{p,q} \frac{\partial G}{\partial u_{p,q}} u_{p,q+1}. \notag
\end{align}
The sums are finite since we will assume that $G$ depends only on finitely 
many derivatives.

\begin{example}
The KdV equation \cite{MAandPCbook1991} for $u(x,t)$ can be recast in 
dimensionless variables as
\begin{equation}
\label{KdV} 
u_t + \alpha u u_x + u_{3x} = 0.
\end{equation}
The parameter $\alpha$ can be scaled to any real number. 
Commonly used values are $\alpha = \pm 1$ or $\alpha = \pm 6.$
A Lax pair for (\ref{KdV}) is given by \cite{Lax1968}
\begin{align}
\LL &= \DD_x^2 + \tfrac{1}{6} \alpha u \, \II,  \notag
\mlabel{KdVL}
\MM &= -4 \DD_x^3 - \alpha u\, \DD_x - \tfrac{1}{2} \alpha u_x\, \II \notag
\end{align}
where $\DD_x^n$ denotes repeated application of $\DD_x$ ($n$ times) and
$\II$ is the identity operator.
Substituting $\LL$ and $\MM$ into (\ref{LaxOpEq}) yields
\begin{align}
\label{KdVLpsi}
\DD_x^2 \psi &= \left( \lambda - \tfrac{1}{6} \alpha u \right) \psi,\\
\label{KdVMpsi}
\DD_t  \psi &= 
- 4 \DD_x^3\psi - \alpha u \, \DD_x\psi - \tfrac{1}{2} \alpha u_x \psi.
\end{align}
The first equation is a Schr\"odinger equation for the eigenfunction 
$\psi$ with eigenvalue $\lambda$ and potential $u(x,t).$ 
The second equation governs the time evolution of the eigenfunction.

The compatibility condition for the above system is
\begin{equation}
\label{directcompatibility}
\DD_t\DD_x^2\psi - \DD_x^2\DD_t\psi = 
\tfrac{1}{6} \alpha  \left( u_t + \alpha u u_x + u_{3x} \right) \psi=0
\end{equation}
where (\ref{KdVLpsi}) and (\ref{KdVMpsi}) are used to eliminate 
$\DD_t\psi,\ \DD_x^2\psi,\ \DD_x^3\psi$ and $\DD_x^5\psi$.
Obviously, \eqref{KdVLpsi} and \eqref{KdVMpsi} will only be compatible on 
solutions of (\ref{KdV}).
\end{example}
\vskip 1pt
\noindent
The compatibility of (\ref{LaxOpEq}) may be expressed directly in terms of 
the operators $\LL$ and $\MM$.
Indeed
\[
\DD_t \left(\LL \psi\right) 
= \LL_t \psi  + \LL \DD_t \psi = \lambda \DD_t \psi 
\]
where $\LL_t \psi \equiv \DD_t \left(\LL \psi \right) - \LL \DD_t \psi$. 
Using \eqref{LaxOpEq} we have
\[  \LL_t \psi + \LL \MM \psi 
= \lambda \MM \psi = \MM \lambda \psi = \MM \LL \psi;\]
that is,
\[
\left(\LL_t + \LL\MM-\MM\LL\right)\psi = 0.
\]
If this operator does not vanish then \eqref{LaxOpEq} is not involutive and 
so would have additional (nonlinear) constraints; 
that is, one cannot freely specify the initial data for \eqref{LaxOpEq}. 
However, if this operator vanishes {\em identically} then Lax pair will not 
encode the original differential equation. 
Therefore, for a non-trivial Lax pair for \eqref{systempde}, 
this operator must vanish only on solutions of \eqref{systempde}.  
Hence
\begin{equation}
\label{opLaxEq}
\LL_t + \left[ \LL, \MM \right] \eqe \opz,
\end{equation}
where $\eqe$ denotes that this equality holds only on solutions to the 
original PDE (\ref{systempde}). 
Here $[\LL, \MM] \equiv \LL \MM - \MM \LL$ is the commutator of the 
operators and $\opz$ is the zero operator. 
Equation (\ref{opLaxEq}) is called the \laxeq. 
Note that $\LL_t = [ \DD_t, \LL]$ and so the Lax equation takes the form
\bq 
[\DD_t - \MM, \LL] \eqe \opz.  \label{geoLaxEq} 
\eq
\begin{example}
Returning to (\ref{KdVL}) for the KdV equation, we have
\begin{align*}
\LL_t  &= [ \DD_t, \LL] = \tfrac{1}{6} \alpha u_t \,\II, \\
\LL\MM &= {} - 4 \,\DD_x^5 - \alpha \Big( \tfrac{5}{3}  u\,\DD_x^3 
 + \tfrac{5}{2}  u_x \,\DD_x^2
 + \left( \tfrac{1}{6} \alpha u^2 + 2 u_{xx} \right)\DD_x
 + \left( \tfrac{1}{12} \alpha u u_x
 + \tfrac{1}{2} u_{3x}  \right)\II \Big), \\
\MM\LL &= {} - 4\,\DD_x^5 - \alpha \Big( \tfrac{5}{3}  u\DD_x^3
 + \tfrac{5}{2} u_x \,\DD_x^2
 + \left( \tfrac{1}{6} \alpha u^2 + 2 u_{xx} \right)\DD_x 
 + \left( \tfrac{1}{4} \alpha u u_x
 + \tfrac{2}{3}  u_{3x} \right)\II \Big).
\end{align*}
Therefore,
\[
\LL_t + [ \LL, \MM] = 
\tfrac{1}{6} \alpha \left(u_t+\alpha u u_x + u_{3x}\right) \II,
\]
which is equivalent to (\ref{directcompatibility}).
\end{example}
\vskip 1pt
\noindent
Various alternatives for the Lax pair operators exist.
For example, one could define $\LH = \LL - \lambda \II $
and $\MH = \MM - \DD_t$ then \eqref{LaxOpEq}  becomes 
$\LH \psi = 0$ and $\MH\psi =0$.
The Lax equation \eqref{opLaxEq} is then $[\LH, \MH]\eqe \opz$.

The order of the operator $\MM$ may also be reduced. 
If the order of $\LL$ is $\ell$ then any terms $\DD_x^{\ell+r}$ in
$\MM$ may be rewritten in terms of derivatives of order at most $\ell -1$. 
Note that this occurs at the expense that the reduced operator 
$\hat{\MM}$ has explicit dependency on the eigenvalue $\lambda$.
\begin{example}
For the KdV equation, using (\ref{KdVL}), we note that 
\bq 
\DD_x^3 \eqe \DD_x \left[\left(\lambda - \tf{6}\alpha u\right) \II\right] 
= \left(\lambda - \tf{6}\alpha u\right) \DD_x - \tf{6} \alpha u_x \, \II  
\label{reducedD} 
\eq
and so the third order operator $\MM$ is equivalent to
\begin{equation}
\label{KdVM1firstorder}
\hat{\MM} = -\left(4 \lambda + \tfrac{1}{3} \alpha  u\right) \DD_x
+ \tfrac{1}{6} \alpha  u_x\,\II,
\end{equation}
which is of first order but depends on $\lambda.$
\end{example}
There is a gauge freedom in the choice of a Lax pair. 
Suppose $\gS$ be an arbitrary but invertible operator and let
${\hat{\psi}} = \gS \psi$.  
Note that 
$\lambda  \hat{\psi} 
= \lambda {\gS}\psi = \gS \LL \psi = \gS \LL \gS[-1] \hat{\psi}$
and
\[ 
\DD_t \hat{\psi} 
= [\DD_t, \gS] \psi + \gS \DD_t\psi 
= \left([\DD_t, \gS] + \gS \MM \right) \gS[-1] \hat{\psi}. 
\]
However, $[\DD_t, \gS]\,\gS[-1]=\DD_t - \gS \DD_t \gS[-1]$ and so 
$\gS \DD_t \gS[-1] \hat{\psi} = \gS \MM \gS[-1] \hat{\psi}$.
Therefore,
\bq
\hat{\LL} = \gS \LL \gS[-1], \qquad
\hat{\MM} = \gS \MM \gS[-1], \label{Opgauge} \qquad
\hat{\DD}_t = \gS \DD_t \gS[-1] 
\eq
will satisfy
$[ \hat{\DD}_t - \hat{\MM}, \hat{\LL} ] \eqe \opz.$

\section{Lax Pairs in Matrix Form}
\label{sec-laxpairsmatrices}
In \cite{MAandJKandANandHS1974}, Ablowitz {\em et al.\/} introduced a matrix 
formalism for Lax pairs. 
Their construction avoids the need to consider higher order Lax operators. 
They associated matrices $\XX$ and $\TT$ to the operators $\LL$ and $\MM$ 
respectively and considered the system
\begin{align}
\DD_x \pvec &= \XX\pvec, \notag
\mlabel{MatrixLax}
\DD_t \pvec & = \TT\pvec \notag 
\end{align}
for an auxiliary vector function $\pvec$.
Both $\XX$ and $\TT$ will be dependent on $\lambda$.
The number of components of $\pvec$ is determined by the order of $\LL$. 
When $\LL$ is of order $2$, $\pvec$ has two components and $\XX$ and  $\TT$ 
are $2 \times 2$ matrices. 
The compatibility condition for \eqref{MatrixLax} is
\[ 
[\DD_t, \DD_x] \pvec = 
\DD_t \left(\XX \pvec\right) - \DD_x\left( \TT \pvec \right) 
= \left(\DD_t  \XX\right) \Psi + \XX \DD_t \pvec- \left(\DD_x \TT\right) \Psi
  - \TT\DD_x\pvec = \zmat; 
\]
that is,
\[ 
\left(\DD_t \XX - \DD_x \TT + \left[ \XX, \TT \right] \right) \pvec = \zmat
\]
where $[\XX, \TT]\equiv \XX \TT - \TT \XX$ is the matrix commutator. 
Since we wish to freely specify initial data, the matrix \laxeq\ follows:
\begin{equation}
\label{matLaxEq}
\DD_t \XX- \DD_x \TT+[\XX,\TT] \eqe \zmat.
\end{equation}
Geometrically, we may interpret \cite{ZakharovFAA79} the matrices $\XX$ and 
$\TT$ as defining a connection on a vector bundle over the base space given 
by $(x, \ t)$. 
$\pvec$ is an element of the vector bundle and parallel transport of this 
element on the base space is given by \eqref{MatrixLax}; that is, 
parallel transport along the $x$-direction is given by
$\left( \DD_x - \XX\right) \Psi = \zmat$
and along the $t$-direction by $ \left( \DD_t - \TT\right) \Psi = \zmat. $
Note that
\[ 
\left[\DD_t -\TT, \DD_x -\XX\right] = \DD_t \XX- \DD_x \TT+[\XX,\TT]. 
\]
Hence, \eqref{matLaxEq} states that the parallel transport of $\pvec$ is 
independent of the path taken.  
For this reason \eqref{matLaxEq} is also known as the zero-curvature equation.
In particular, we may rewrite \eqref{matLaxEq} as 
$[ \cov[t], \cov] \eqe \zmat$ where $\cov[t] \equiv \DD_t - \TT$ and $\cov
\equiv\DD_x - \XX$ are the components of the {\em covariant derivative} 
induced by the connection.
\begin{example}
It is well-known that
\begin{equation}
\label{KdVX}
\XX =
\left(
\begin{array}{cc}
0 & \quad 1\\
\lambda-\tfrac{1}{6} \alpha u & \quad 0
\end{array}
\right)
\end{equation}
and
\begin{equation}
\label{KdVT}
\TT =
\left(
\begin{array}{cc}
\tfrac{1}{6} \alpha u_x 
& \quad - 4 \lambda - \tfrac{1}{3} \alpha u
\\ 
- 4\lambda^2 + \tfrac{1}{3} \alpha \lambda u + \tfrac{1}{18} \alpha^2  u^2
+ \tfrac{1}{6} \alpha u_{xx} 
& \quad {} -\tfrac{1}{6} \alpha u_x
\end{array}
\right)
\end{equation}
%
%
form a Lax pair for the KdV equation since
\[
\DD_t \XX- \DD_x \TT+[\XX, \TT]
= -\tfrac{1}{6} \alpha 
\left(
\begin{array}{cc}
 0 & \quad 0 \\
 u_t + \alpha u u_x + u_{3x} & \quad 0
\end{array}
\right)
\]
which evaluates to the zero matrix on solutions of (\ref{KdV}).
\end{example}
Given an operator \lax, $(\LL,\MM)$, finding a corresponding matrix
\lax, $(\XX,\TT)$, is a straightforward but lengthy computation.
As noted above in \eqref{reducedD}, the operator $\LL$ allows one to write 
higher order derivatives in $x$ in terms of lower order derivatives. 
Let the order of $\LL$ be $\ell$. 
Without loss of generality, we may assume that the leading coefficient of 
$\LL$ is $1$. 
Thus,
\bq \LL =  
\DD_x^{\ell}  + f_{\ell -1}\, \DD_x^{\ell -1} + \cdots + f_0 \,\II. 
\label{Ll} \eq
Let
\[ \pvec = \mat[c]{\psi \\ \DD_x \psi \\ \vdots \\ \DD_x^{\ell -1} \psi} \]
and so
\bq \setlength{\arraycolsep}{8pt}\DD_x \pvec 
= \mat[c]{\DD_x \psi \\ \DD_x^2 \psi \\ \vdots \\ \DD_x^{\ell } \psi} 
= \mat[ccccc]{0 & 1 & 0 & \cdots & 0 \\ 
0 & 0 & 1 & \cdots & 0 \\ 
\vdots & \vdots & \vdots & \ddots & \vdots \\ 
0 & 0 & 0 & \cdots & 1 \\ 
\lambda- f_0 & {}- f_1 & {} - f_2 & \cdots & - f_{\ell-1}} \pvec 
= \XX \pvec. \label{L2XX} \eq 
The computation of $\TT$ is more elaborate. 
We have
\[ \DD_t \pvec = \mat[c]{\psi_t \\ \DD_x \psi_t \\ 
\vdots \\ 
\DD_x^{\ell-1} \psi_t} = \mat[c]{\MM \psi \\ 
\DD_x \MM \psi \\ 
\vdots \\ 
\DD_x^{\ell-1} \MM \psi}. \]
Note that, using \eqref{LaxOpEq} and \eqref{Ll},
\bq\DD_x^{\ell + r} \psi 
= {}- \DD_x^{r} \left(  f_{\ell -1} \,\DD_x^{\ell -1} 
  + \cdots + f_1\, \DD_x + (f_0 -\lambda)\,\II \right) \psi. 
\label{higherrx} \eq
Therefore any derivatives of $\psi$ in $x$ of order $\ell$ or higher are 
removed by applying \eqref{higherrx} recursively. 
The result is a vector which will depend linearly on $\psi$ and its first 
$\ell -1$ derivatives; that is 
%
$ \DD_t \pvec = \TT \pvec$.

We will only show the conversion explicitly for the KdV equation.
\begin{example}
For the KdV equation, \eqref{L2XX} reduces to \eqref{KdVX}.
Using \eqref{KdVM1firstorder}, we note that
\begin{align*}
 \DD_t \psi &= \tfrac{1}{6} \alpha u_x\psi
  - \left(4\lambda + \tfrac{1}{3} \alpha u\right) \DD_x \psi, \\
 \DD_t\DD_x\psi &= \tfrac{1}{6} \alpha u_{xx}\,\psi -\tfrac{1}{6} \alpha u_x\,
 \DD_x \psi  
  - \left(4\lambda + \tfrac{1}{3} \alpha u\right) \DD_x^2 \psi.  
\end{align*}
However,
$ \DD_x^2 \psi = \left(\lambda - \tf{6}\alpha u \right) \psi $
and so
\[
\DD_t\DD_x\psi =\left(-4\lambda^2 + \tfrac{1}{3} \alpha \lambda u
 + \tfrac{1}{18} \alpha^2 u^2 + \tfrac{1}{6} \alpha u_{xx} \right) \psi
 - \tfrac{1}{6} \alpha u_x\,\DD_x \psi .  \]
Therefore,
\[
\DD_t \pvec = 
\left(
\begin{array}{cc}
\dsp \;\;\tfrac{1}{6} \alpha u_x & \quad
\dsp - 4\lambda - \tfrac{1}{3} \alpha u 
\\
\dsp -4\lambda^2 + \tfrac{1}{3} \alpha \lambda u
 + \tfrac{1}{18} \alpha^2 u^2
 + \tfrac{1}{6} \alpha u_{xx}
 & \quad \dsp  - \tfrac{1}{6} \alpha u_x
\end{array}
\right) \pvec 
\]
which yields $\TT$ in (\ref{KdVT}).
\end{example}
There is gauge freedom in the construction of the matrix representation 
of a Lax pair.  
Let $\myG$ be a non-singular matrix and $\hat{\pvec} = \myG \pvec$. 
Using \eqref{MatrixLax},
\[ \DD_x \hat{\pvec} 
= \myG \DD_x\pvec + \DD_x \left(\myG\right) \pvec 
= \left(\myG \XX + \DD_x \left(\myG \right)\right) \pvec
= \left(\myG \XX + \DD_x\left(\myG\right)\right) \myG^{-1} \hat{\pvec} 
= \hat{\XX} \hat{\pvec} \]
where
\bq 
\hat{\XX} 
= \myG\XX\myG^{-1} +\DD_x \left(\myG \right) \myG^{-1}. 
\label{gaugeX}
\eq
Similarly, $\DD_t \hat{\pvec}=\hat{\TT} \pvec$ with
\bq 
\hat{\TT} = \myG\TT\myG^{-1} +\DD_t \left(\myG\right)\myG^{-1}. 
\label{gaugeT} 
\eq
\begin{example}
A second \lax\ \cite{MAandPCbook1991} for the \kdv\ equation consists of 
the complex matrices
\begin{equation}
\label{KdVX1}
\XXH = 
\left( \begin{array}{cc}
-i k & \quad \dsp \tfrac{1}{6} \alpha u \\ 
-1   & \quad  i k
\end{array} \right)
\end{equation}
and
%
%
\begin{equation}
\label{KdVT1}
\TTH = 
\left(
\begin{array}{cc}
-4 i k^3 + \tfrac{1}{3} i \alpha  k u - \tfrac{1}{6} \alpha u_x 
& \quad \tfrac{1}{3} \alpha \left( 2 k^2 u - \tfrac{1}{6} \alpha u^2
+ i k u_x - \tfrac{1}{2} u_{xx} \right) \\
\;\; - 4 k^2 + \tfrac{1}{3} \alpha  u 
& \quad \;\; 4 i k^3 - \tfrac{1}{3} i \alpha k u + \tfrac{1}{6} \alpha  u_x
\end{array}
\right).
\end{equation}
Based on (\ref{gaugeX}) and (\ref{gaugeT}), both \lax{s}\ are gauge equivalent
with 
\[
\myG =
\left( \begin{array}{cc}
- i k & \quad 1 \\ 
- 1 & \quad 0
\end{array} \right)
\]
where $\lambda = -k^2.$
\end{example}

\section{Dilation Invariance of Nonlinear Evolution Equations}
\label{dilationinvariancePDEs}
 
Crucial to the computation of Lax pairs is that (\ref{opLaxEq}) must hold 
on the PDE (\ref{systempde}). 
This has the important consequence that any symmetry of the PDE,
in particular the dilation symmetry, might be useful to find (\ref{opLaxEq}). 
In other words, we make an ansatz that the operators $\LL$ and $\MM$ inherit 
the scaling symmetry from the PDE. 
This, as it turns out, greatly simplifies  the equations that determine the 
Lax pair.
\begin{example}
The KdV equation (\ref{KdV}) is {\em dilation invariant} under the 
scaling symmetry
%
\begin{equation}
\label{KdVscale} 
(t, x, u) \rightarrow ({\kappa}^{-3} t,\kappa^{-1} x, {\kappa}^{2} u),
\end{equation}
where $\kappa$ is an arbitrary parameter. 
Indeed, after a change of variables with 
$\tilde{t} = {\kappa}^{-3} t, \tilde{x} 
= {\kappa}^{-1} x, \tilde{u} = {\kappa}^{2} u$ 
and cancellation of a common factor $\kappa^{-5},$ the KdV for 
$\tilde{u}(\tilde{x},\tilde{t})$ arises. 
This dilation symmetry can be expressed as $u \sim \partial_x^2$
and $\partial_t \sim \partial_x^3$ which means that $u$ scales as 
two $x-$derivatives and the $t$-derivative scales as three $x-$derivatives.
\end{example}
We define the {\em weight}, $W,$ of a variable as the exponent of $\kappa$ 
in the scaling symmetry. 
It is clear that, from \eqref{TotalD}, $\DD_x \sim \partial_x$ and 
$\DD_t \sim \partial_t$ and so we extend weights to operators. 
For (\ref{KdVscale}),  $W(x) = -1$ (or $W(\partial_x) = 1), 
W(t) = -3$ (or $W(\partial_t) = 3$) and $W(u) = 2.$ 
However we could have scaled each of these weights (effectively 
$\kappa \rightarrow \kappa^s$); that is, a scaling symmetry does {\em not} 
uniquely determine a set of weights. 
We use this freedom to set $W(\partial_x)=1$.

The weight of a monomial is defined as the {\em total} weight 
of the monomial and will also be denoted by $W$. 
Such monomials may involve the independent and dependent variables and the 
operators $\partial_x,\ \DD_x,\ \partial_t$ and $\DD_t.$ 
In particular, note that $W(\DD_x)=W(\partial_x)$ and $W(\DD_t)=W(\partial_t)$.
An expression (or equation) is {\em uniform in weight} if its monomial terms 
have equal weights. 
For example, (\ref{KdV}) is uniform in weight since each of the three terms 
has weight $5.$

The importance of uniformity of weight is demonstrated by the KdV equation 
\eqref{KdV}. From (\ref{KdVL}), it is clear that $\LL$ has weight 2 and 
$\MM$ has weight 3 since $W(\II)=0$.
Therefore this Lax pair inherits the scaling symmetry. 
Also note that we must assign $W(\lambda)=2$. 
The elements of the matrices $\XX$ and $\TT$ in (\ref{KdVX}) and (\ref{KdVT}) 
are also uniform in weight, albeit of different weights.  
This is also true for the matrices in (\ref{KdVX1}) and (\ref{KdVT1}).
%
%
\section{An algorithm for Computing Lax Pairs in Operator Form}
\label{sec-algorithm}
First, one computes the dilation symmetry of the nonlinear PDE and assigns 
weights such that $W(\partial_x)=1$.  
Next, one selects the weight $\ell$ of the operator $\LL$. 
Since we have chosen $W(\partial_x)=1$, $\ell$ will also be the order 
of $\LL$.  
The minimal weight for $\LL$ is the maximum weight of the dependent 
variables since we want the Lax equation to depend {\em non-trivially} on 
the PDE. 
It is only through $\LL_t$ that the $t$-derivatives of the dependent variables
appear. 
Since $\LL_t=[\DD_t,\LL]$, $W(\LL_t) = W(\LL) + W(\partial_t)$ and, 
from \eqref{opLaxEq}, $W(\LL_t) = W(\LL) + W(\MM)$. 
Thus $W(\MM)=W(\partial_t)$. Since we require that $\MM$ to be a 
{\em differential} operator, we must choose a system of weights such that 
$W(\partial_t)$ is a non-negative integer.

Thus the candidate Lax pair has the form
\begin{align*}
\LL &=  \DD_x^{\ell}  + f_{\ell -1} \,\DD_x^{\ell -1} + \cdots + f_0 \,\II, \\
\MM &= c_m \, \DD_x^m + g_{m-1} \, \DD_x^{m-1} + \cdots + g_0 \,\II
\end{align*}
where $m=W(\partial_t)$.  
Now
\bq
\left[\LL, \MM\right] =\left(\ell \DD_x c_m \right) \DD_x^{m+\ell-1} + \cdots. 
\label{ExpandLM} 
\eq
Recall that we require that  \eqref{opLaxEq} only holds on solutions of 
the PDE. 
Thus the operator $\DD_x^{m+\ell-1}$ could be reduced to an operator that 
depends on derivatives of order strictly less than $\ell$. 
However this would introduce the eigenvalue $\lambda$ and, in this case, 
there would be a term $\ell \lambda^{m-1} \DD_x c_m$. 
This would be the {\em only} term in $\lambda^{m-1}$. 
Under the assumption that the operator $\LL$ has a complete (and therefore 
infinite) set of eigenvalues, we conclude $\DD_x c_m = 0;$ that is, 
$c_m$ is a {\em constant}. 
Therefore  \eqref{ExpandLM} becomes
\[ \left[\LL, \MM\right] 
= \left( \ell \DD_x g_{m-1} - m c_m \DD_x f_{\ell-1} \right) \DD_x^{m+\ell-2} 
+ \cdots. \]
After reduction, the coefficient of $\lambda^{m-2}$ must vanish and so
\bq 
\ell \DD_x g_{m-1} - m c_m \DD_x f_{\ell-1}  = 0. \label{kinematic1} 
\eq
In this way, we obtain $m-1$ equations for the unknown coefficient functions 
$f_i, \ g_j$ that do {\em not} depend on the PDE. 
We call these equations {\em kinematic constraints}.

The remaining $\ell$ components of \eqref{opLaxEq} {\em may} involve 
terms from $\LL_t$. 
When $t$-derivatives {\em do} appear, we must use the PDE to remove them. 
We call these equations {\em dynamical equations}.  

The kinematic constraints are easily solved. 
All have the form  \eqref{kinematic1} and so may be solved in terms of the 
leading $g_j$. 
Specifically, \eqref{kinematic1} yields
\[  
g_{m-1} =  \tf[m]{\ell} c_m f_{l-1}. 
\]
In this way, $g_j$, $j=1, \ \ldots, \ m-1$, may be written in terms of $g_0$ 
and $f_i$. 
Note that this process does {\em not} depend on the PDE and so this rewrite 
is {\em canonical}.

After the elimination of the $g_j$, the dynamical equations are a set of 
ODEs for the remaining coefficient functions $g_0$ and $f_i$. 
We now make the ansatz that the terms in the operators $\LL$ and $\MM$ 
have uniform weight; that is, $f_i$ has weight $\ell-i$ and $g_0$ 
has weight $m$. 
Thus, for example, $g_0$ is assumed to be a linear combination 
(with unknown coefficients) of monomials of weight $m$. 
If, in addition, the weights of the dependent variables are positive and 
the monomials do {\em not} depend explicitly on $x$ and $t$, the dynamical 
equations reduce to an {\em finite} system of {\em algebraic} equations for 
the unknown coefficients. 
This algebraic system (which will also have any parameters that are present 
in the PDE) is then solved by Gr\"{o}bner basis methods.

We now illustrate the steps of the algorithm for the KdV equation.
\subsection{Step 1: Computing the scaling symmetry}
\label{sec-scalingsymmetry}
The dilation symmetry of (\ref{KdV}) can be readily computed by the 
requirement that \eqref{KdV} is uniform in weight.  
Indeed, setting $W(\partial_x) = 1$, we have
\[
W(u) + W(\partial_t) = 2 W(u) + 1 = W(u) + 3,
\]
which yields $W(u) = 2$ and $W(\partial_t) = 3$. 
This confirms (\ref{KdVscale}). 
So, the requirement of {\em uniformity in weight} of a PDE allows one to 
compute the weights of the variables (and thus the scaling symmetry) with 
linear algebra.

Dilation symmetries, which are special Lie-point symmetries, are common 
to many nonlinear PDEs. 
Needless to say, not every PDE is dilation invariant. 
However non-uniform PDEs can be made uniform by extending the set of 
dependent variables with auxiliary parameters with appropriate weights. 
Upon completion of the computations one can set these parameters to $1$.
\subsection{Step 2: Building a candidate Lax pair}
\label{sec-candidatelaxpair}
Since $W(u)=2$, the minimal weight for $\LL$ is $2$. 
We therefore choose $\ell=2$. 
Note, that there is no monomial of weight $1$ 
(any monomial that involves $u$ must have weight at least $2$). 
Therefore we {\em could} set $f_1=0$. 
However the kinematic constraints and dynamical equations {\em only} depend on
the choice of weight for $\LL$ and $W(\partial_t)$.  
So, by not setting $f_1=0$ at this stage, we will derive the dynamical 
equations for {\em all} Lax pairs with $\LL$ of second order and 
$W(\partial_t)=3$. 
Our candidate Lax pair is
\begin{align} \LL &= \DD_x^2 +f_1 \,\DD_x +  f_0 \,\II,  \notag
\mlabel{23LaxPair}
\MM &= c_3\, \DD_x^3 +g_2 \,\DD_x^2+ g_1\, \DD_x + g_0 \,\II. \notag
\end{align}
The kinematic constraints (that is, the coefficients of $\DD_x^3$ and 
$\DD_x^4$ in $[\LL,\MM]$) give
%
%
\[
g_2 = \tf[3]{2} c_3 f_1,  \qquad
g_1 = \tf[3]{8} c_3 \left( 2 \DD_x f_1 + f_1^2 + 4 f_0 \right).
%
\]
Therefore, \eqref{23LaxPair} must have the form
\begin{align} \LL &= \DD_x^2 +f_1 \,\DD_x +  f_0 \,\II,  \notag
\mlabel{23LaxPairred}
\MM &= 
c_3\, \DD_x^3 +\tf[3]{2} c_3 f_1 \,\DD_x^2
+ \tf[3]{8}c_3\left( 2 \DD_x f_1 + f_1^2 + 4f_0 \right) \DD_x + g_0 \,\II. 
\notag
\end{align}
The dynamical equations are
\begin{align}
\DD_t f_1 + 2\DD_x g_0 - \tf{8} c_3 \DD_x 
\left( 2 \DD_x^2 f_1 + 12 \DD_x f_0 -  f_1^3 + 12 f_1f_0 \right) &=0, \notag\\
\DD_t f_0 +  \DD_x^2 g_0  + f_1 \DD_x g_0 \hspace*{7.5cm} & \label{23LaxEq} \\
{} - c_3 \left( \DD_x^3 f_0 + \tf[3]{2} f_1 \DD_x^2 f_0 
+ \tf[3]{4} \DD_x f_1 \DD_x f_0 + \tf[3]{8} f_1^2 \DD_x f_0 
+ \tf[3]{2}  f_0 \DD_x f_0 \right)&=0.  \notag
\end{align}
Equations \eqref{23LaxEq} are the master equations for \eqref{23LaxPairred}. 
In other words, for {{\em any} equation with a scaling symmetry such that 
$W(\partial_t)=3$, a Lax pair with a second order eigenvalue problem must 
have the form \eqref{23LaxPairred} where $f_0, \ f_1, \ g_0$ and $c_3$ 
satisfy \eqref{23LaxEq}.

For the case of the KdV equation, we have $f_1=0$ and so \eqref{23LaxEq} 
reduces to
\begin{align*}
 2\DD_x g_0 -   \tf[3]{2} c_3\DD_x^2 f_0 &=0, \\
\DD_t f_0 +  \DD_x^2 g_0 - c_3 \left( \DD_x^3 f_0 + \tf[3]{2}
 f_0 \DD_x f_0 \right)&=0.
\end{align*}
Note that, in this case, the first dynamical equation does not depend on the 
PDE and so may be regarded as a kinematic constraint. 
This also implies that we {\em cannot} set $c_3=0$ in order to satisfy the 
kinematic constraint since this would result in a trivial dynamical equation 
$\DD_t f_0 =0$.
Therefore we have $g_0 = \tf[3]{4} c_3 \DD_x f_0$ and
\bq 
\DD_t f_0 - \tf{4} c_3 \DD_x^3 f_0  - \tf[3]{2} c_3 f_0 \DD_x f_0 =0. 
\label{KdVd0} 
\eq
Since $f_0$ has weight $2$ we must have
\bq 
f_0 = b_0 u \label{KdVf0} 
\eq 
with $b_0 \neq 0$. 
\subsection{Step 3: Computing the undetermined coefficients}
\label{sec-undeterminedcoeffs}
Equation \eqref{KdVf0} is substituted into the dynamical equation 
\eqref{KdVd0}.  
We obtain
\[
b_0 u_t  - c_3b_0 u_{3x} - \tf[3]{2} c_3 b_0^2 u u_x  =0 .
\]
Use (\ref{KdV}) to replace $u_t$ by $-(\alpha u u_x + u_{3x})$ and so
\[ 
b_0 \left(\alpha u u_x + u_{3x} + \tf{4} c_3u_{3x} 
+ \tf[3]{2} c_3b_0 u u_x \right) =0. 
\]
Equating to $0$ the coefficients of the monomials $u_{3x}$ and $uu_x$, 
we have
\[
b_0 \left(1 + \tf{4}c_3\right) =0, \qquad
b_0 \left( \alpha + \tf[3]{2} c_3 b_0\right) =0.
\]
Since $b_0 \neq 0$, we must have $c_3 =  -4$ and $b_0 = \tf{6} \alpha$.
Finally substitute $f_0= \tf{6} \alpha u, \ f_1=0, \ g_0=-\tf{2}
\alpha u_x$ and $c_3=-4$ into \eqref{23LaxPairred} to obtain the Lax pair 
\eqref{KdVL}.  
In this case the algebraic system was sufficiently simple that it did not 
require the machinery of Gr\"{o}bner bases to solve.
%
\section{Examples}
\label{sec-examples}
The method described in Section~\ref{sec-algorithm} can be applied to 
many nonlinear PDEs in $(1+1)$ dimensions with polynomial nonlinearities. 
In this section we apply the method to scalar equations as well as systems. 
The featured scalar examples are the modified KdV and Boussinesq equations.
The method is also illustrated for systems, including a coupled system of KdV 
equations due to Hirota \& Satsuma and the Drinfel'd-Sokolov-Wilson system.
%
\subsection{The modified Korteweg-de Vries equation}
\label{ex-mkdv}
For the modified Korteweg-de Vries (mKdV) equation \cite{MAandPCbook1991} 
for $u(x,t)$,
\begin{equation}
\label{mKdV}
u_t + \alpha {u}^2 u_x + u_{3x} = 0,
\end{equation}
the weights may be chosen to be $W(u)=1, W(\partial_x)=1,$ and 
$W(\partial_t)=3.$
Therefore the minimal weight for $\LL$ is $1$ and the weight of $\MM$ is $3$.

Suppose $\LL$ has order $1$; that is, 
\begin{align}
\LL &= \DD_x + f_0 \, \II, \notag
\mlabel{mKdV13}
\MM &= c_3\, \DD_x^3 + g_2 \,\DD_x^2 + g_1 \,\DD_x + g_0 \, \II.  \notag
\end{align}
The kinematic constraints are
\[
\DD_x  g_2 -3c_3 \, \DD_x f_0 = 0, \qquad
\DD_x g_1 - 3c_3 \DD_x^2 f_0 - 2g_2\DD_x f_0  =0
\]
and so
\[
g_2 = 3 c_3 f_0, \qquad
g_1 = 3 c_3 ( \DD_x f_0 + f_0^2 ).
\]
The dynamical equation is
\[ 
\DD_t f_0 + \DD_x g_0 -c_3 \DD_x \left( \DD_x^2 f_0 - 3f_0\DD_x f_0 
- f_0^3  \right) = 0. 
\]
In this case, we can replace $g_0$ by
$ g_0 - c_3 \left( \DD_x^2 f_0 -3f_0\DD_x f_0 - f_0^3  \right) $
to remove $c_3$. Thus we can set $c_3=0$ without loss of generality. 
Therefore the Lax pair \eqref{mKdV13} is
\begin{align}
\LL &= \DD_x + f_0 \, \II, \notag
\mlabel{mKdV13red}
\MM &= g_0 \, \II \notag
\end{align}
with the dynamical (that is, master) equation
\[ \DD_t f_0 + \DD_x g_0 = 0.\]

Now assume that $f_0$ has weight $1$ and $g_0$ has weight $3$; that is 
\[
f_0 = b_0 u, \qquad
g_0 = a_1 u^3 + a_2 uu_x + a_3 u_{xx}
\]
with $b_0 \neq 0$. The dynamical equation becomes
\[ 
{} - b_0 \left( \alpha u^2u_x + u_{3x}\right) 
+ 3a_1 u^2u_x + a_2 \left(u_x^2 +uu_{xx}\right) + a_3 u_{3x}   = 0. 
\]
Setting to zero the coefficients of the monomials $u_{3x}, \ uu_{xx}, 
\ u_x^2$ and $u^2u_x$ yields
\[
a_3 - b_0 = 0, \qquad a_2 =0, \qquad 
3a_1 - b_0 \alpha =0 . 
\]
Since $b_0$ is free, we can set $b_0=1$ and \eqref{mKdV13red} becomes
\begin{align}
\LL_1 &= \DD_x + u \,\II,  \notag
\mlabel{mKdV_LM_1}
\MM_1 &=   \left(\tf{3}\alpha u^3 + u_{xx} \right) \II. \notag
\end{align}
The Lax equation for \eqref{mKdV_LM_1} is the conservation law form 
of \eqref{mKdV}.

If we assume that $\mathcal{L}$  has weight $2$, then, since
$W(\partial_t)=3$, the Lax pair is given by \eqref{23LaxPairred}
and the dynamical equations are given by \eqref{23LaxEq}. 
In this case, a gauge on $g_0$ will not remove $c_3$ from the dynamical 
equations.  
The ansatz of uniform weights determines
\[
f_1 = b_0 u, \qquad
f_0 = b_1 u^2 + b_2 u_x, \qquad
g_0 = a_1 u^3 + a_2 uu_x + a_3 u_{xx}.
\]
The solution to resultant algebraic system has two branches;
\begin{align*}
& \text{Case I:}&
\LL &= \DD_x^2 + 2u \,\DD_x + \left( u^2 + u_x\right) \II, \\
&& \MM &= \left(  \tf{3}\alpha u^3 + u_{xx} \right) \II. \\
& \text{Case II:}&
\LL &= \DD_x^2 + 2\epsilon u \, \DD_x + \tf{6} \left( 6\epsilon^2u^2 
+6\epsilon u_x+\alpha u^2 \pm \sqrt{-6\alpha}u_x\right) \II, \\
&&\MM &= {} -4 \,\DD_x^3 -12 \epsilon u \, \DD_x^2 
- \left( 12 \epsilon^2 u^2 + 12 \epsilon u_x 
+ \alpha u^2 \pm \sqrt{-6\alpha}u_x\right) \DD_x \\
&&& \qquad {}-  
\left( 4 \epsilon^3u^3+\tf[2]{3} \epsilon \alpha u^3+12\epsilon^2 u u_x 
\pm \epsilon \sqrt{-6\alpha} uu_x +3 \epsilon u_{xx} \right. \\
&&& \qquad \left. {} +\alpha u u_x \pm \tf{2}\sqrt{-6\alpha}u_{xx}\right) \II.
\end{align*}
The second case is a $1$-parameter family (with parameter $\epsilon$) of 
Lax pairs. 
This family has the property $\MM \rightarrow {} -4 \DD_x^3$ for solutions 
of \eqref{mKdV} such that $u \rightarrow 0$ as $|x| \rightarrow \infty$. 
Wadati \cite{MW1972,MW1973} used this property to construct soliton solutions 
for \eqref{mKdV} via the inverse scattering transform from Case II with 
$\epsilon=0$.

There is one Lax pair with $\LL$ of order $3$ which is
\begin{align*}
\LL 
&= \DD_x^3 + 3u \,\DD_x^2 +3 \left(u^2+u_x\right) \DD_x 
+ \left( u^3 + 3uu_x + u_{xx} \right) \II, \\
\MM &=\left(  \tf{3} \alpha u^3 + u_{xx} \right) \II.
\end{align*}
However
\[ \LL = (\DD_x + u\II)^3 = \LL_1^3, \qquad \MM = \MM_1 \]
where $\LL_1$ and $\MM_1$ are given by \eqref{mKdV_LM_1}.
%
\subsection{The Boussinesq equation}
\label{ex-boussinesq}
The wave equation,
\begin{equation}
\label{boussinesq}
u_{tt} - u_{xx} + 3 u_x^2 + 3 u u_{xx} + \alpha u_{4x} = 0,
\end{equation}
for $u(x,t)$ with real parameter $\alpha,$ was proposed by Boussinesq 
\cite{Bous1872} to describe surface waves in shallow water 
\cite{MAandPCbook1991}. 
This equation may be rewritten as a system
\begin{equation}
\label{boussinesqsys} 
u_t - v_x = 0, \quad 
v_t - u_x + 3 u u_x + \alpha u_{3x} = 0,
\end{equation}
where $v(x,t)$ is an auxiliary dependent variable.
We could proceed following the above examples modulo the discussion on 
scaling below. 
However we will work directly with \eqref{boussinesq}. 
First note that  \eqref{boussinesq} can be written as a conservation law,
\bq 
\DD_t u_{t} = \DD_x \left( u_x - 3 u u_x - \alpha u_{3x} \right). 
\label{bouscon} 
\eq
However it is not uniform in weight (the monomials $u_{x}$ and
$uu_{x}$ cannot have the same weight unless $W(u)=0$). 
This problem may be circumvented by the introduction of an auxiliary
parameter $\beta$ with an associated weight. 
Thus, we replace
\eqref{bouscon} by
\bq 
\DD_t u_{t} = \DD_x \left(\beta  u_x - 3 uu_x 
- \alpha u_{3x} \right) \label{bousbeta} 
\eq
and assign weights
\[ 
W(\partial_x) = 1, \ W(\partial_t) = 2, \ W(u)=2 \ \text{and}
\ W(\beta)=2. 
\] 
This example demonstrates that a PDE, which is not dilation invariant, 
may be expressed in a dilation invariant form by the introduction of 
auxiliary parameters with weights.
Upon completion of the computations, one can set each of these parameters 
to $1.$

The ``fundamental" dependent variable in this equation is $u_t$
(that is, $u_{tt}=\DD_t u_t$). 
However the Boussinesq equation is a conservation law. 
Therefore we can use $\DD_x^{-1} u_t$ as a fundamental variable; in other 
words, rewrite \eqref{bousbeta} as 
\[ 
\DD_t \left(\DD_x^{-1} u_t \right)
= \beta  u_x - 3 uu_x - \alpha u_{3x}. 
\]
Therefore the minimal order of $\LL$ is $3$. 
Since there are no weight $1$ monomials, the minimal order Lax pair is given by
\begin{align*}
\LL &= \DD_x^3 + f_1 \DD_x + f_0 \II, \\
\MM &= c_2 \DD_x^2 + g_0 \II.
\end{align*}
The kinematic constraint is $3\DD_x g_0 = 2 c_2 \DD_x f_1 $ which gives
\[ g_0 = \tf[2]{3} c_2 f_1 + c_0 \beta. \]
The extra term arises since the constant $\beta$ has weight $2$.
The dynamical equations become
\begin{align}
\DD_t f_1 &=c_2 \left( 2 \DD_x f_0 - \DD_x^2 f_1\right), 
\label{bousdyn_1}  \\
\DD_t f_0 &= c_2 \left( \DD_x^2 f_0 -\tf[2]{3} \DD_x^3 f_1 
- \tf[2]{3} f_1 \DD_x f_1 \right).  \notag
\end{align}
Since it is $u_{tt}$ and {\em not} $u_t$ that appears in \eqref{bousbeta}, 
derivatives of $u$ that have at most only one $t$-derivative may be freely 
specified in the initial data for \eqref{bousbeta}. 
Thus $u_t, \ u_{xt}, \ u_{xxt}, \ \ldots$ may appear in monomials used to 
construct $f_0$ and $f_1$. 
In addition, $f_0$ {\em only} appears in the right hand side of the
dynamical equations in the form $\DD_x f_0$. 
Thus a {\em non-local} term of the form $\DD_x^{-1} f$, where $f$ 
has weight $4$,may appear in $f_0$. Consequently, the weight ansatz implies
\begin{align*}
f_1 &=a_1 u + a_2\beta, \\
f_0 
&= a_3 u_x + \DD_x^{-1} \left( a_4 u^2 +a_5 \beta u + a_6 u_t 
+ a_7 \beta^2\right).
\end{align*}
Under this ansatz, the dynamical equation \eqref{bousdyn_1} becomes the 
kinematic constraint
\bq 
a_1 u_t = 2c_2 \left( a_3 u_{xx} + a_4 u^2+ a_5 \beta u + a_6 u_t 
+ a_7 \beta^2 - \tf{2} a_1 u_{xx} \right).  
\eq
For a non-trivial solution, we require $a_1 \neq 0$ and so $c_2 \neq 0$. 
Thus $a_4=a_5=a_7=0$, $a_1=2c_2a_6$ and $a_3=c_2a_6$.
Hence,
\[ 
\DD_t f_0 = a_3 u_{xt} + a_6 \DD_x^{-1} u_{tt} \eqe  a_6 \left( c_2 u_{xt} 
+ \beta u_x -3uu_x - \alpha u_{3x} \right) 
\]
on solutions of \eqref{bousbeta}. 
Therefore the remaining dynamical equation is
\[ 
a_6\left( c_2 u_{xt} + \beta u_x -3uu_x - \alpha u_{3x} \right) 
=  c_2 a_6\big( u_{xt}  - \tf{3} c_2 u_{3x}  
- \tf[4]{3} c_2 u_x(2c_2a_6u+a_2\beta) \big). 
\]
The solution of this equation gives the Lax pair
\begin{align*}
\LL &= \DD_x^3 + \frac{1}{4\alpha}\left( 3u - \beta\right) \DD_x 
       + \frac{3}{8\alpha^2} \left(\alpha u_x 
       \pm \tf{3}\sqrt{3\alpha} \, \DD_x^{-1} u_t \right) \II, \\
\MM &={} \pm \sqrt{3\alpha}\,\DD_x^2 \pm \frac{\sqrt{3\alpha}}{2\alpha} u\,\II.
\end{align*}
The subcase $\beta=1$ is a Lax pair for \eqref{boussinesq} \cite{VZ1974}. 
Note that the auxiliary variable $v$ that was introduced to obtain the 
system \eqref{boussinesqsys} is, in fact, $\DD_x^{-1} u_t$. 
A Lax pair for the system is obtained by replacing $\DD_x^{-1} u_t$ by $v$. 
%
\subsection{The coupled Korteweg-de Vries equations}
\label{ex-hssystem}
The coupled Korteweg-de Vries (cKdV) equations \cite{RHandJS1981}, 
\begin{equation}
\label{ckdv}
u_t - 6 \beta u u_x + 6 v v_x - \beta u_{3x} = 0, 
\quad 
v_t + 3 u v_x + v_{3x} = 0
\end{equation}
where $\beta$ is a nonzero parameter, describes interactions of two waves 
with different dispersion relations. 
System (\ref{ckdv}) is known in the literature as the Hirota-Satsuma system.
It is completely integrable \cite{MAandPCbook1991,RHandJS1981}
when $\beta = \frac{1}{2}.$

We assign the weights $W(\partial_x)=1, \ W(\partial_t) = 3, \ W(u) = 2$ 
and $W(v) = 2$. 
Since $W(\partial_t)=3$ and there are no weight $1$ monomials, the dynamical 
equations are precisely those of the KdV equation \eqref{KdV}.
The minimal weight for $\LL$ is $2$ and so the dynamical equation is 
\eqref{KdVd0} with 
\[ 
f_0 = b_0u + b_1v. 
\]
The resultant equations have only a trivial solution. 
Likewise, there are also no Lax pairs with $\LL$ of third order.

For the case $W(\LL)=4$, the candidate Lax pair, after the kinematic 
constraints have been solved, is
\begin{align*}
\LL &= \DD_x^4 + f_2\, \DD_x^2 + f_1\, \DD_x + f_0 \,\II, \\
\MM &= c_3\, \DD_x^3 + \tf[3]{4} c_3 f_2 \,\DD_x 
+ \tf[3]{8} c_3\left(2f_1 -\DD_x f_2\right) \II
\end{align*}
with dynamical equations
\begin{align*}
\DD_t f_2 +\tf{4}c_3\left(6\DD_x^2 f_1 
- \DD_x^3f_2+3f_2\DD_xf_2-12\DD_x f_0\right)&=0,\\
\DD_t f_1 +\tf{4}c_3\left(8\DD_x^3 f_1 -3\DD_x^4f_2+3f_2\DD_xf_1 
+3f_1\DD_xf_2-12\DD_x^2f_0\right)&=0, \\
\DD_t f_0 + \tf{8}c_3\left(6\DD_x^4f_1-3\DD_x^5f_2+6f_2\DD_x^2f_1 
-3f_2\DD_x^3f_2 \right. \hspace*{2cm} &\\
\left. {}+6f_1\DD_x f_1-3f_1\DD_x^2f_1-8\DD_x^3f_0-6f_2\DD_xf_0\right) &=0.
\end{align*}
A non-trivial solution exists only when $\beta=\tf{2}$. 
In this case the solution is
%
%
\begin{align*}
\LL &=  \DD_{x}^4+2 u \,\DD_{x}^2+2 \left(u_{x}-v_x\right) \DD_{x}
+\left(u^2-v^2+u_{2x}-v_{2x}\right) \II, \\ 
\MM &= 2\,\DD_{x}^3+3 u\, \DD_{x} +\tf[3]{2}\left(u_{x} - 2 v_{x} \right) \II.
\end{align*}
Note that this Lax pair is given in \cite{RDandAF1982,GWpla1982}.
%
\subsection{The Drinfel'd-Sokolov-Wilson equations}
\label{ex-drinfeld}
We consider a one-parameter family of the Drinfel'd-Sokolov-Wilson (DSW) 
equations
\begin{equation}
\label{DSWsys} 
u_t + 3 v v_x = 0, \quad 
v_t + 2 u v_x + \alpha u_x v + 2 v_{3x} = 0,
\end{equation}
where $\alpha$ is a nonzero parameter. 
The system with $\alpha = 1$ was first proposed by Drinfel'd and Sokolov
\cite{VDandVSsmd1981,VDandVSjsm1984} and Wilson \cite{GWpla1982}. 
It can be obtained \cite{JimboRIMS83} as a reduction of the 
Kadomtsev-Petviashvili equation (a two-dimensional version of the KdV 
equation).

We may assign weights $W(\partial_t)=3$, $W(u)=2$ and $W(v)=2$.
Thus we have the same weight system as the KdV equation and the coupled 
system above. 
There are no Lax pairs for this system with $\LL$ of order $2$, $3$, $4$ 
or $5$ which inherit this scaling symmetry.

There are no solutions for $W(\LL)=6$ except for $\alpha=1$. 
In that case 
%
%
\begin{align*}
\LL &= \DD_x^6 + 2 u \, \DD_x^4 + \left(4 u_x-3 v_x\right) \DD_x^3 +
\left(\tf[9]{2} \left (u_{2x}-v_{2x}\right)+u^2-v^2\right) \DD_x^2 \\
 & \qquad {}+\left(\tf[5]{2} \left(u_{3x}-v_{3x}\right)
+2 \left( u u_x-v v_x \right)+u_x v-u v_x\right) \DD_x \\
& \qquad {} + \tf{4} \left( 2 \left (u_{4x} - v_{4x} \right ) 
+ 2 \left( u+v \right) \left ( u_{2x}-v_{2x} \right)
+ u_x^2 - v_x^2 \right) 
\II, \\
\MM &= \DD_x^3+u \, \DD_x-\tf{2} \left(3 v_x-u_x\right) \II.
\end{align*}
The DSW system (\ref{DSWsys}) is known to have infinitely many conservation
laws when $\alpha=1$ and is completely integrable 
\cite{VDandVSsmd1981,GWpla1982}. 
This is the Lax pair given in \cite{GWpla1982}. 
%
\section{Classification of fifth-order KdV-like equations}
For an equation or system with parameters, the algorithm described in 
Section~\ref{sec-algorithm} can be used to find the necessary conditions on 
the parameters so that the PDE has a Lax pair and therefore may be completely
integrable.

Consider the family of fifth-order KdV-type equations with three parameters
\bq u_t + \alpha u^2 u_x + \beta u_x u_{xx} + \gamma u u_{3x} + u_{5x} = 0. 
\label{KdV5} \eq
This family includes several well-known completely integrable equations. 

Replacing $u$ by $\frac{u}{\gamma}$, it is clear that only the ratios 
$\frac{\alpha^2}{\gamma}$ and $\frac{\beta}{\gamma}$ matter. 
We continue with (\ref{KdV5}) for an easier comparison with the results 
in the literature. 
The special cases of (\ref{KdV5}) shown in the table are extensively 
discussed in the literature 
\cite{UGandWH1997,RHandMI1983,BKandGW1981,JSandDK1977}. 
Only the first three equations are completely integrable. 
Ito's equation is not completely integrable but has an unusual set of
conservation laws \cite{UGandWH1997,MI1980}.
\begin{center}
\begin{table}[] 
\begin{tabular}{ | c | c | c | c |} 
\hline 
& & & \\
\parbox{2cm}{\centering{Parameter ratios\\[1mm] $ 
\dsp (\frac{\alpha}{\gamma^2}, \frac{\beta}{\gamma})$ }} 
& \parbox{2cm}{\centering Commonly used values\\[1mm] $ 
\dsp (\alpha,\beta,\gamma)$}& Equation name 
& \parbox{2cm}{\centering Lax pair references} \\
&&& \\
\hline
& & & \\
$\dsp (\tfrac{3}{10},2)$ & $\dsp (30,20,10),(120,40,20),$ & Lax 
& \cite{Lax1968}\\ 
& $(270,60,30)$ &&\\
& & & \\
$\dsp (\tfrac{1}{5},1)$ & $ \dsp (5,5,5),(180,30,30),$ 
& Sawada-Kotera 
& \cite{RDandJG1977,FordyJMP80}\\
& $(45,15,15)$ && \cite{DK1980,JSandDK1977} \\
& & & \\ 
$\dsp (\tfrac{1}{5},\tfrac{5}{2})$ & $\dsp (20,25,10)$ & Kaup-Kupershmidt
& \cite{FordyJMP80,DK1980} \\ 
& & & \\ 
$\dsp (\tfrac{2}{9},2)$ & $\dsp (2,6,3),(72,36,18)$ & Ito  
&  --\\ 
& & & \\
\hline
\end{tabular}
\vskip 5pt
\begin{center} 
Special cases of the fifth-order family (\ref{KdV5}) of KdV-type equations. 
\end{center}
\end{table}
\end{center}
We wish to determine which members, if any, have a  Lax pair, with $\LL$ of 
second order, which may then be amenable to a (standard) inverse scattering 
transform based on a quadratic eigenvalue problem. 
Kaup \cite{DK1980} has considered this family with respect to a {\em cubic} 
eigenvalue problem (that is $\LL$ is a third order operator).

Uniformity of weights implies that $W(\partial_t) = 5$. 
The first kinematic constraint is 
\[ 
2 \DD_x g_4 - 5 c_5 \DD_x f_1 = 0 .
\]
In this situation we {\em may need}  or {\em wish} to give the parameters 
weight. 
If this is the case, there may be a weight $1$ constant.
Therefore the solution to this constraint is
\[ 
g_4 = \tf[5]{2} c_5 f_1 + c_4 
\]
where $c_4$ is a constant of integration of weight $1$. 
After all four kinematic constraints are solved, we have
\begin{align*}
\LL &=  \DD_x^2 + f_1\, \DD_x + f_0\, \II, \\
\MM &= c_5 \,\DD_x^5 + \left(\tf[5]{2} f_1 + c_4\right) \DD_x^4 
+ \Big(\tf[5]{8} c_5(6\DD_x f_1 +3f_1^2+4f_0)+2c_4f_1+c_3 \Big) \DD_x^3 \\
&\quad {} + \Big(\tf[5]{16} c_5 (12\DD_x^2 f_1+10 \DD_x f_1^2+12 \DD_x f_0
+f_1^3+12f_1f_0) +c_4(2\DD_xf_1+f_1^2+2f_0) \\
&\quad  {}+\tf[3]{2}c_3f_1+c_2 \Big)\DD_x^2 
+ \Big(\tf[5]{128}c_5(48f_0\DD_xf_1+96f_1\DD_xf_0+4\DD_xf_1^3+24f_1^2f_0 \\
&\quad  {}+20(\DD_x f_1)^2+\left. 40f_1\DD_x^2f_1
+24\DD_x^3f_1+80\DD_x^2f_0+48f_0^2-f_1^4) \right. \\ 
&\quad {}+c_4(2\DD_x f_0+f_1\DD_x f_1+\left. 
\DD_x^2f_1+2f_1f_0)+\tf[3]{8}c_3(4f_0+f_1^2+2\DD_xf_1) \right. \\
&\quad {} +c_2f_1+c_1\Big) \DD_x +g_0\, \II.
\end{align*}
For brevity, we omit the two dynamical equations.

We now need to make a choice for the remaining weights. 
The obvious choice is $W(u)=2$. 
With this choice, $\alpha, \ \beta$ and $\gamma$ all have weight zero. 
Therefore the constants of integration, $c_1,  \ldots, c_4$, are all zero 
which simplifies the Lax pair dramatically. 
The only non-trivial solution occurs when 
$\alpha = \tf[3]{10} \gamma^2, \ \beta = 2\gamma$ 
which corresponds to the fifth-order KdV equation due to Lax \cite{Lax1968}. 
The Lax pair is
\begin{align*}
\LL &= \DD_x^2 + \tf{10} \gamma u \,\II, \\
\MM &= {} -16 \, \DD_x^5 - 4\gamma u\,  \DD_x^3 -6\gamma u_x \, \DD_x^2  
- \gamma\left(5u_{xx} + \tf[3]{10}\gamma u^2 \right)\DD_x 
- \gamma\left(\tf[3]{2} u_{3x} + \tf[3]{10}\gamma u u_x \right) \II. 
\end{align*}
In \cite{CalogeroJMP91}, it is shown that all scalar evolution equations, 
which have the form of a  conservation law, have a Lax pair with a second 
order $\LL$. 
Such Lax pairs can be found if we choose the weights 
$W(u)=1, \ W(\alpha)=2, \ W(\beta) =1$ and $W(\gamma)=1$. 
In this case, the constants of integration may not be zero.

For weight $3$, we obtain non-trivial solutions for the Sawada-Kotera and 
Kaup-Kupershmidt equations. 
For the Sawada-Kotera equation ($\alpha=\tf{5} \gamma^2, \\beta=\gamma$) 
\cite{PCandRDandJG1976,KSandTK1974} there are two solutions: 
%
%
\begin{align*} &\text{Case I:}&  
\LL &= \DD_{x}^3+\tfrac{1}{5}\gamma u\, \DD_{x},
\\
&& \MM &= 9 \, \DD_{x}^5+3 \gamma u\, \DD_{x}^3+3\gamma u_{x} \,\DD_{x}^2
+ \gamma \left( \tfrac{1}{5}\gamma u^2 + 2 u_{2x}\right) \DD_{x} \\
&\text{Case II:}&
\LL &= \DD_{x}^3+\tfrac{1}{5}\gamma u\, \DD_{x}+\tfrac{1}{5}\gamma u_{x}\, 
\mathcal{I},
\\
&&\MM &= 9\, \DD_{x}^5+3\gamma u\, \DD_{x}^3+6\gamma u_{x}\, \DD_{x}^2
+ \gamma \left(\tfrac{1}{5}\gamma u^2 + 5  u_{2x}\right)\DD_{x} \\
&&& \qquad\qquad {} 
+ 2 \gamma \left(\tfrac{1}{5}\gamma uu_{x} + u_{3x}\right)\mathcal{I}.
\end{align*}
The first Lax pair is given in \cite{RDandJG1977,FordyJMP80}.

The Kaup-Kupershmidt equation ($\alpha=\tf{5}\gamma^2, \
\beta=\tf[5]{2} \gamma$)  \cite{DK1980} has a single Lax pair
%
%
\begin{align*}
\LL &= \DD_{x}^3+\tfrac{1}{5}\gamma u\,\DD_{x}+\tfrac{1}{10}\gamma u_{x}\, 
\mathcal{I}, 
\\ 
\MM &= 9\,\DD_{x}^5+3\gamma u\, \DD_{x}^3+\tfrac{9}{2}\gamma u_{x} \, 
\DD_{x}^2
+\gamma \left(\tfrac{1}{5}\gamma u^2+\tfrac{7}{2} u_{2x}\right)\DD_{x} 
+\gamma \left(\tfrac{1}{5}\gamma uu_{x} + u_{3x}\right)\mathcal{I}
\end{align*}
which is given in \cite{FordyJMP80}. 
We cannot find any reference to the second Lax pair in the literature.
%
\section{Conclusions}
\label{sec-conclusions}
The algorithm described in this paper can be easily implemented in 
{\sc Mathematica} or {\sc Maple}. 
Therein lies its value; it gives a quick method to ``test the waters" with 
a new system of equations. 
If the equations do not have a scaling symmetry, or the scaling symmetry does 
not yield a Lax pair, then weighted parameters may be introduced. 
A new assignment of weights may lead to a Lax pair. 
The ``focus on the equation" of our approach is a marked contrast to existing 
work where integrable systems have, by and large, been developed from 
geometric considerations (for example, the use of the Drinfel'd--Sokolov 
construction \cite{VDandVSsmd1981,VDandVSjsm1984}).

The question of an appropriate system of weights is non-trivial.
As noted above, all members of the three parameter family of fifth--order 
KdV--like equations have a Lax pair with a second order $\LL$. 
The question naturally arises as to which of these Lax pairs are useful. 
Can, for example, a criteria be developed that would guarantee an infinite 
family of local conservations laws?
\section*{Dedication}
This paper is dedicated to the memory of Alan Jeffrey.  
As a Ph.D.\ student, one of us (WH) was first introduced to the mathematics 
of nonlinear waves through Alan's seminal paper on 
Nonlinear Wave Propagation published in 
Zeitschrift f\"ur Angewandte Mechanik und Mathematik {\bf 58}, T38-56 (1978).
With Alan's passing we loose a researcher, educator, and expositor par 
excellence.
%
\section*{Acknowledgements}
\label{sec-acknowledgements}
This material is based in part upon research supported by the
National Science Foundation (NSF) under Grant No.\ CCF-0830783.
Any opinions, findings, and conclusions or recommendations expressed in 
this material are those of the authors and do not necessarily reflect the 
views of NSF.

WH is grateful for the hospitality and support of the Department of Computer 
Engineering at Turgut \"{O}zal University (Ke\c{c}i\"{o}ren, Ankara, Turkey) 
where code for Lax pair computations was further developed.

MH would like the thank the Department of Applied Mathematics and Statistics, 
Colorado School of Mines for their hospitality while this work was completed.

Undergraduate students Oscar Aguilar, Sara Clifton, William ``Tony" McCollom, 
and graduate student Jacob Rezac are thanked for their help with this project.
%

\end{document}